\documentclass[showpacs,preprintnumbers,amsmath,amssymb,prb,twocolumn,floatfix,amssymb]{revtex4}
\usepackage[dvips]{graphicx,color}
\usepackage{dcolumn}
\usepackage{amsmath,amssymb,bbold,bm}

\begin{document}
\title{Magnetoelectric Response of the Time-Reversal Invariant Helical Metal 
}
\author{Ion Garate$^{1,2}$ and M. Franz$^{1}$}
\affiliation{$^1$Department of Physics and Astronomy, The University of British Columbia, Vancouver, BC V6T 1Z1, Canada}
\affiliation{$^2$Canadian Institute for Advanced Research, Toronto, ON M5G 1Z8, Canada.}
\date{\today}

\begin{abstract}
We derive compact analytical expressions for the coupled spin-charge susceptibility of a clean helical metal at the surface of a three dimensional topological insulator (TI). 
These expressions lead to unconventional non-collinear RKKY interactions between two impurity magnetic moments placed on the surface of a TI, and predict the generation of electric currents by time-dependent magnetic moments.
We determine the influence of gate and bias voltages on the interlayer exchange coupling between two single-domain ferromagnetic monolayers deposited on top of a TI.
\end{abstract}
\maketitle

{\em Introduction.---}
The discovery of three dimensional topological insulators\cite{ti} has unveiled a promising avenue for low-dissipation spintronics applications.
The surfaces of these materials host topologically protected, gapless, strongly spin-orbit (SO) coupled conducting states dubbed helical metals.
These are endowed with novel magnetoelectric properties when time-reversal (${\cal T}$) symmetry is broken \cite{qi2008,yokoyama2009,garate2009}.
Albeit less exotic, the magnetoelectric response of ${\cal T}$-invariant helical metals is also peculiar and potentially useful.
Recent theoretical studies\cite{ye2010,gao2009,biswas2009,liu2009} have  discussed the spin-spin reponse of ${\cal T}$-invariant helical metals, with an emphasis on RKKY interactions.
This paper complements and extends previous work by presenting fully analytical expressions for the spin-spin as well as spin-charge response of a clean surface.
In anticipation of future experiments we predict that (i) the helical metal mediates Dzyaloshinskii-Moriya coupling when the chemical potential lies away from the neutral point, (ii) electric fields  may be used to generate and tune unconventional RKKY interactions; (iii) precessing magnetic moments produce alternating electric currents. 

{\em Spin-charge response function.---}
The low-energy effective Hamiltonian describing a helical metal living on the surface of a three dimensional TI is\cite{ti}
\begin{equation}
\label{eq:h0}
{\cal H}_0=-i v\int d^2{\bf r}\sum_{\alpha\beta}\Psi_\alpha^\dagger ({\bf r}){\boldsymbol\sigma}_{\alpha\beta}\cdot(\hat{z}\times{\boldsymbol\nabla})\Psi_{\beta} ({\bf r}),
\end{equation}
where $v$ is the Fermi velocity, ${\boldsymbol\sigma}=(\sigma^x,\sigma^y,\sigma^z)$ are Pauli matrices denoting real spin, $\alpha,\beta\in\{\uparrow,\downarrow\}$, ${\bf r}=(x,y)$, $\hat z$ is a unit vector normal to the surface and $\hbar\equiv 1$.
The density matrix encoding the ground state charge and spin densities is
\begin{equation}
\rho_a({\bf r},{\bf r}')=\sum_{\alpha\beta}\langle\Psi_\alpha^\dagger({\bf r})\sigma^a_{\alpha\beta} \Psi_{\beta}({\bf r}')\rangle,
\end{equation}
where $a\in\{0,x,y,z\}$ labels charge ($0$) or spin ($x,y,z$) sectors and $\sigma^0\equiv{\bf 1}$ is the identity matrix.
Under a weak external potential 
\begin{equation}
U^{\rm ext}_{\alpha\beta}=\sum_{a} U_a\sigma^a_{\alpha\beta}
\end{equation}
of frequency $\Omega$ and wave vector ${\bf q}=q(\cos\phi,\sin\phi)$, the density matrix changes via
\begin{equation}
\delta\rho_a ({\bf q},\Omega)=\chi_{ab}({\bf q},\Omega) U_b ({\bf q},\Omega)
\end{equation}
where
\begin{equation}
\label{eq:chi_ab}
\chi_{ab}({\bf q},\Omega)=-i\int_{{\bf k},\omega}
{\rm Tr}\left[\sigma^a G({\bf k},\omega)\sigma^b G({\bf k+q},\omega+\Omega)\right]
\end{equation}
is the dynamical spin-charge susceptibility and 
\begin{eqnarray}
\label{eq:green}
&&G({\bf k},\omega)= i\frac{-\omega {\bf 1}+v{\boldsymbol\sigma}\cdot({\hat z\times\bf k})}{-\omega^2+v^2 k^2-i 0^+}\nonumber\\
&&+\pi\frac{-\omega {\bf 1}+v{\boldsymbol\sigma}\cdot(\hat z\times {\bf k})}{v k}\delta(\omega-v k)\Theta(\mu-v k)
\end{eqnarray}
is the Green's function for the surface states.\cite{barlas2007} $\Theta$ is the step function and $\mu= v k_F$ is the chemical potential.
From Eqs.~(\ref{eq:chi_ab}) and ~(\ref{eq:green}) we arrive at
\begin{equation}
\label{eq:chi_sc}
\chi_{ab}=\left(\begin{array}{cccc} \chi_{00}         &  - f_{0}\sin\phi             &   f_{0}\cos\phi           &  0\\
                                   -f_{0}\sin\phi     &    f_{1} \cos^2\phi          &  \frac{f_{1}}{2}\sin2\phi    & -i f_{2} \cos\phi\\
                                    f_{0}\cos\phi     &    \frac{f_{1}}{2}\sin2\phi   &  f_{1} \sin^2\phi          & -i f_{2} \sin\phi\\
                                     0                  &   i f_{2} \cos\phi            &  i f_{2} \sin\phi              &  f_{3}
\end{array}\right),
\end{equation}
the form of which is compatible with symmetry arguments discussed in the context of ordinary two-dimensional electron systems (2DES) with Rashba SO interactions\cite{huang2006}.
The charge-charge response function $\chi_{00}$ is identical to that of graphene\cite{graphene} and there is no coupling between the charge density and the $z$-component of the spin density.
Adopting the formalism explained in Ref.~[\onlinecite{barlas2007}] we have derived the following concise expressions for the coefficients $\{f_i\}$:
\begin{eqnarray}
\label{eq:f_123}
&&f_1(q,0) =\frac{-q}{16 v}+\frac{q}{8\pi v}{\rm Re}\left[x\sqrt{1-x^2}+\sin^{-1}{x}\right]\nonumber\\
&&f_2(q,0)=\frac{q}{4\pi v}\left[1-{\rm Re}\sqrt{1-x^2}\right] \\
&&f_3(q,0)=\frac{-q}{8 v}-\frac{k_F}{2\pi v}+\frac{q}{4\pi v}{\rm Re}\left[\sin^{-1}{x}\right] \nonumber
\end{eqnarray}
with $x=2k_F/q$ and
\begin{eqnarray}
\label{eq:f_0}
f_0(q,\Omega)&=&\frac{-1}{16 v}\frac{q\,\Omega}{\sqrt{q^2 v^2-\Omega^2}}+\frac{\Omega\, k_F}{2\pi q v^2}\\
&-&\frac{q\, \Omega}{8 \pi v\sqrt{q^2 v^2-\Omega^2}}{\rm Re}\left[\sin^{-1}{y}+y\sqrt{1-y^2}\right], \nonumber
\end{eqnarray}
with $y=(2k_F+\Omega/v)/q$.
We omitted $\Omega\neq 0$ expressions for $f_{1,2,3}$; they are cumbersome and will not be needed below.
$f_1(0,0)=f_2(0,0)=0$ indicates that, unlike in ordinary 2DES with SO interactions\cite{imamura2004}, a uniform and static {\em in-plane} magnetic field does not spin-polarize the helical metal: it merely shifts the location of the Dirac cone in momentum space\cite{caveat}. 
In contrast, uniform and static perpendicular magnetic fields elicit a paramagnetic response $\propto f_3(0,0)\neq 0$.
Finally, $f_{0}({\bf q},0)=0$ means that charge and spin sectors decouple in the static limit.

{\em RKKY interaction in electric equilibrium ---}
Consider two localized spins ${\bf S}_1$ and ${\bf S}_2$ placed at ${\bf R}_1$ and ${\bf R}_2$ on the surface of a 3D TI. 
Their indirect RKKY\cite{rkky} coupling mediated by the helical metal is 
\begin{equation}
E_{\rm RKKY}=\sum_{i,j} J_{ij} S_1^i S_2^j,
\end{equation}
where $i,j\in\{x,y,z\}$, $J_{ij}=-\lambda_{ij}^2 \chi_{ij}({\bf R})$, ${\bf R}={\bf R}_1-{\bf R}_2$ and $\chi_{ij}({\bf R})$ is the spatial Fourier transform of $\chi_{ij}({\bf q},0)$. In addition
$\lambda_{ij}$ are the exchange integrals (in units of energy $\times$ area) between the magnetic impurities and the surface states. Invoking spin rotational invariance in the $xy$ plane we have $\lambda_{xx}=\lambda_{yy}=\lambda_{xy}=\lambda_{yx}\equiv \lambda_{||}$,  $\lambda_{zz}\equiv\lambda_{\perp}$ and $\lambda_{xz}=\lambda_{zx}=\lambda_{yz}=\lambda_{zy}=\sqrt{\lambda_{||}\lambda_{\perp}}$.

\begin{figure}[h]
\begin{center}
\includegraphics[scale=0.3]{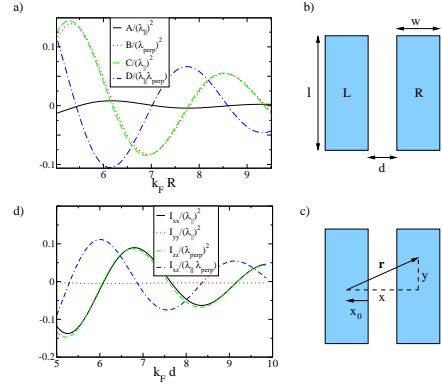}
\caption{\label{fig:fig1} (a) RKKY coefficients between two impurity spins in the $k_F R\gg 1$ regime (Eq.~(\ref{eq:large kf})). We take $k_F^3/(16\pi^2 v)\equiv 1$. At large distances $A$ is parametrically smaller than $B,C,D$. (b) Two ultrathin ferromagnetic strips deposited on the surface of a 3D TI (viewed from the top). (c) The integration variables in Eq.~(\ref{eq:I}). (d) Interlayer exchange coefficients $I_{ij}$ for two uniform ferromagnetic monolayers, as a function of the interlayer separation. We took $16 \pi^2 v s_L s_R\equiv 1$, $k_F w=25$, $k_F l=1000$.}
\end{center}
\end{figure}

Combining Eqs.~(\ref{eq:chi_sc}) and (\ref{eq:f_123}) we arrive at
\begin{eqnarray}
E_{\rm RKKY}&=&A (S_1^x S_2^x+S_1^y S_2^y)+B S_1^z S_2^z\\
            &+&C({\bf S}_1\cdot\hat{\bf R})({\bf S}_2\cdot\hat{\bf R})
            +D[\hat{\bf R}\times({\bf S}_1\times{\bf S}_2)]_z\nonumber
\end{eqnarray}
which agrees with the results derived independently in Ref.~[\onlinecite{ye2010}]. 
We are able to extract analytical expressions for the RKKY coefficients in limiting cases:
\begin{equation}
\label{eq:kf_0}
A =\frac{\lambda_{||}^2}{32 \pi v R^3} \mbox{   ,   } B =\frac{-\lambda_{\perp}^2}{16\pi v R^3} \mbox{   ,   } C = -3 A \mbox{   ,   } D=0
\end{equation}
for $k_F=0$ and
\begin{eqnarray}
\label{eq:large kf}
A &=&\frac{\lambda_{||}^2}{8\pi^2 v R^3}\cos(2 k_F R)\\
B &=&\frac{-\lambda_{\perp}^2}{16\pi^2 v R^3}\left[3 \cos(2 k_F R)+4 k_F R \sin(2 k_F R)\right]\nonumber\\
C &=&\frac{-\lambda_{||}^2}{16 \pi^2 v R^3}\left[5 \cos(2 k_F R)+4 k_F R \sin(2 k_F R)\right]\nonumber\\
D &=&\frac{-\lambda_{||} \lambda_{\perp}} {16\pi^2 v R^3}\left[4 k_F R \cos(2 k_F R)-3 \sin(2 k_F R)\right] \nonumber
\end{eqnarray}
for $k_F R\gg 1$ (see Fig.\ref{fig:fig1}a).
In Eq.~(\ref{eq:kf_0}) we have followed the prescription of Ref.~[\onlinecite{saremi2007}] to regularize ultraviolet divergences.
In the derivation of Eq.~(\ref{eq:large kf}) we have integrated by parts and have used asymptotic expansions for integrals involving polynomials of Bessel functions.   
When $k_F=0$ the RKKY interaction decays monotonically as $1/R^3$.
When $k_F R\gg 1$ all the coefficients have oscillatory behavior and the leading terms decay as $1/R^2$; this
behavior is characteristic of an ordinary 2DES\cite{huang2006,imamura2004}, except for the fact that $B,C,D$ depend on $k_F$. 
Density-dependent amplitudes are ubiquitous in graphene\cite{brey2007} as well, where nonetheless $C=D=0$.
Finally, we verify that $(A+C)/\lambda_{||}^2=B/\lambda_{\perp}^2$ for any $k_F$, which is also satisfied in a 2DES with Rashba SO coupling\cite{imamura2004}.

We now consider two {\em uniformly magnetized}\cite{baltensperger1990} ferromagnetic monolayers depicted in Fig.~\ref{fig:fig1}b,c.
The interlayer exchange coupling per unit length is
\begin{equation}
{\cal E}_{\rm xc}=\sum_{i,j\in\{x,y,z\}}I_{ij}\Omega_L^i \Omega_R^j,
\end{equation}
where $\hat\Omega_{L (R)}$ is the direction of magnetization in the left (right) ferromagnet,
\begin{equation}
\label{eq:I}
I_{ij}=\frac{-\lambda_{i,j}^2}{s_L s_R}\int_0^w dx_0\int_{x_0+d}^{x_0+d+w} dx \int_{-l/2}^{l/2} d y\,\, \chi_{ij}({\bf r}),
\end{equation}
and $s_{L(R)}$ is the area per spin in the left (right) magnet.
When $k_F=0$ and $l\gg w\gg d$ we find a ferromagnetic interlayer coupling with strong in-plane anisotropy: 
\begin{eqnarray}
\label{eq:E_ex0}
&&{\cal E}_{\rm xc}(\mu=0)= I_{xx} \Omega_L^x\Omega_R^x+ I_{zz}\Omega_L^z\Omega_R^z;\nonumber\\
&&I_{xx}\simeq\frac{-\lambda_{||}^2}{16 \pi v s_L s_R}\log\frac{w}{2 d} \mbox{   ;   } I_{zz}=2\frac{\lambda_{\perp}^2}{\lambda_{||}^2} I_{xx}.
\end{eqnarray}
For $k_F R\gg 1$ we evaluate $I_{ij}$ numerically and arrive at
\begin{equation}
\label{eq:Ex}
{\cal E}_{\rm xc}=I_{xx}\Omega_L^x\Omega_R^x+I_{yy}\Omega_L^y\Omega_R^y+I_{zz}\Omega_L^z\Omega_R^z+I_{zx}\hat y\cdot\left(\hat\Omega_L\times\hat\Omega_R\right)
\end{equation}
with the coefficients displayed in Fig.~\ref{fig:fig1}d.
Once again there is a strong in-plane exchange anisotropy ($I_{\rm yy}\simeq 0$).
 The last term of Eq.~(\ref{eq:Ex}) is a Dzyaloshinskii-Moriya (DM) interaction\cite{dm} that favors non-collinear magnetization configurations. 
The DM vector ${\bf D}=I_{zx}\hat y$ alternates sign as a function of the separation between the ferromagnets and may be tuned by a gate voltage.

{\em RKKY interaction in presence of an in-plane electric field ---}
It is well-known that the amplitude of interlayer exchange coupling between two magnets changes when the non-magnetic spacer is driven out of electric equilibrium\cite{neq}. 
It is less known that electric fields can also induce unconventional, non-Heisenberg types of exchange coupling, provided that the spacer contains spin-orbit interactions.
Due to its strong SO interaction, the surface of a 3D TI constitutes an ideal spacer where such an effect might be observable.

\begin{figure}[b]
\begin{center}
\includegraphics[scale=0.35]{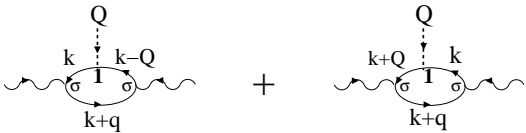}
\caption{\label{fig:fig2} Influence of an in-plane electric field on the RKKY interaction: Feynman diagrams for $\delta\chi_{ij}$. We pick the contribution linear in ${\bf Q}$, which is associated with the response to a uniform in-plane electric field.}
\end{center}
\end{figure}

We compute the change in the static spin-spin susceptibility under a chemical potential $\mu({\bf r})$ which is constant in time and varies slowly in space (Fig.~\ref{fig:fig2}) :
\begin{equation}
\label{eq:chi_E}
\delta\chi_{ij}({\bf q},0; {\bf Q},0)={\bf Q}\cdot{\boldsymbol\Pi}_{ij}\mu({\bf Q},0)+O(Q^2),
\end{equation}
where ${\bf Q}$ is the wave vector of the chemical potential and ${\boldsymbol \Pi}_{ij}$ is given by the ${\bf Q}\to 0$ limit of  
\begin{equation}
\label{eq:Pi}
2\frac{\partial}{\partial {\bf Q}}\int_{{\bf k},\omega}{\rm Tr}\left[\sigma^i G({\bf k}_+,\omega) \sigma^j G({\bf k}_-,\omega) {\bf 1} G({\bf k},\omega)\right],
\end{equation}
${\bf k}_+\equiv{\bf k+q}$, ${\bf k}_-\equiv{\bf k-Q}$ and the factor of $2$ stems from summing the second diagram in Fig.~\ref{fig:fig2}.
For simplicity we assume that the unperturbed system has the chemical potential at the Dirac point; in such case it is possible to evaluate Eq.~(\ref{eq:Pi}) analytically. 
We obtain 
\begin{equation}
\delta\chi_{ij}=\delta\chi_{xz}(\delta_{ix}\delta_{jz}+\delta_{iz}\delta_{jx})+\delta\chi_{yz}(\delta_{iy}\delta_{jz}+\delta_{iz}\delta_{jy}),
\end{equation}
where $\delta_{ij}$ is the Kronecker delta,
\begin{eqnarray}
\label{eq:d_chi_E}
\delta\chi_{xz}({\bf q};{\bf Q})&=&-\frac{i Q \mu}{16 v^2 q}\cos\phi\cos(\phi-\phi_Q)\nonumber\\
\delta\chi_{yz}({\bf q};{\bf Q})&=&-\frac{i Q \mu}{16 v^2 q}\sin\phi\cos(\phi-\phi_Q)
\end{eqnarray}
and we have used ${\bf Q}=Q(\cos\phi_Q,\sin\phi_Q)$.
Fourier transforming from ${\bf q}$ to ${\bf R}$ ,
we see that the in-plane electric field produces non-collinear RKKY interactions between two magnetic impurities:
\begin{eqnarray}
\label{eq:d_Erkky}
\delta E_{\rm RKKY}&=&\sum_{ij} (\delta J_{ij}) S_1^i S_2^j\\
&=&\lambda_{||}\lambda_{\perp}\frac{(\hat{\bf R}\times{e\bf E})_z}{32\pi v^2 R}\left[(\hat{\bf R}\times{\bf S}_1)^z {\bf S}_2^z + 1\leftrightarrow 2\right],
\nonumber
\end{eqnarray}
where $\delta J_{ij}=-\lambda_{ij}^2\delta\chi_{ij}({\bf R})$ and ${\bf E}=-i{\bf Q}\mu/e$ is the electric field.
$\delta E_{\rm RKKY}=0$ when ${\bf E}||{\bf R}$. 
The amplitude of $\delta E_{\rm RKKY}$ equals that of $E_{\rm RKKY}$ (Eq.~(\ref{eq:kf_0})) when ${\bf E}\perp{\bf R}$ and $E\sim E_c\equiv\hbar v/(e R^2)$.
For $v=5\times 10^5 {\rm m/s}$, $E_c[{\rm mV/nm}]\sim 330/R[{\rm nm}]^2$.
Thus moderate electric fields suffice to modify the equilibrium RKKY interaction substantially, especially when the impurities are separated by a distance $\gtrsim 5 {\rm nm}$. 

For the system shown in Fig.~\ref{fig:fig1}b, the change in the interlayer exchange coupling due to an in-plane electric field is
\begin{eqnarray}
\label{eq:d_Eex}
&&\delta{\cal E}_{\rm xc}=\delta I_{xz}\Omega_L^x\Omega_R^z + \delta I_{yz}\Omega_L^y\Omega_R^z + L\leftrightarrow R;\nonumber\\
&&\delta I_{yz}\simeq\frac{-\lambda_{||}\lambda_{\perp} e E_y}{16 \pi v^2 s_L s_R}w^2 \mbox{   ,   } \frac{\delta I_{xz}}{\delta I_{yz}}\simeq\frac{E_x}{E_y}\log\frac{l}{4 w},  
\end{eqnarray}
where we have assumed $l\gg w\gg d$. 
$\delta{\cal E}_{\rm xc}=0$ when $\Omega_L^z=\Omega_R^z=0$ and $\delta{\cal E}_{\rm xc}$ may be nonzero even when $E_x=0\neq E_y$, i.e. when there is no potential difference between the ferromagnets.
Moreover, the amplitude of $\delta{\cal E}_{\rm xc}$ becomes comparable to that of ${\cal E}_{\rm xc}$ (Eq.~(\ref{eq:E_ex0})) when $E\gtrsim\hbar v/ e w^2$.
The task of evaluating the $k_F\neq 0$ counterparts of Eqs.~(\ref{eq:d_Erkky}) and (\ref{eq:d_Eex}) is cumbersome and will be left for future work.

{\em Coupled spin-charge dynamics---}
Eq.~(\ref{eq:f_0}) implies that a time-dependent Zeeman field ${\bf U}$ will induce a charge density $n_{\rm ind}$ on a ${\cal T}$-invariant helical metal.
For $k_F=0$ we find
\begin{equation}
n_{\rm ind}^{(0)}({\bf q},\Omega)=-\frac{e\, \Omega}{16 v\sqrt{v^2 q^2-\Omega^2}}(\hat z\times {\bf q})\cdot {\bf U}({\bf q},\Omega),
\end{equation}
and thus
\begin{equation}
n_{\rm ind}^{(0)}({\bf r},\Omega)=i\frac{e \Omega}{16 v}(\hat z\times\nabla)\cdot\int\frac{d^2{\bf q}}{(2\pi)^2}\frac{{\bf U}({\bf q},\Omega) e^{i {\bf q}\cdot{\bf r}}}{\sqrt{v^2 q^2-\Omega^2}},
\end{equation}
which is valid for an arbitrary Zeeman field. 
Motivated by the possibility of experiments with single spins deposited on the surface of a 3D TI we concentrate on a spatially localized perturbation, whereby $U({\bf q},\Omega)=U(\Omega)$ and
\begin{equation}
n_{\rm ind}^{(0)}({\bf r},\Omega)=i\frac{e \Omega}{16 v}{\bf U}(\Omega)\cdot(\hat z\times\nabla)\left[\frac{\exp\left(i\frac{\Omega r}{v}\right)}{v r}\right].
\end{equation}
It follows that
\begin{equation}
\label{eq:n0}
n_{\rm ind}^{(0)}({\bf r},t)=-\frac{e}{16 v^2}(\hat z\times\nabla)\cdot\left[\frac{1}{r}\frac{\partial{\bf U}(t-r/v)}{\partial t}\right].
\end{equation} 
The induced current defined via $\nabla\cdot{\bf j}_{\rm ind}=-\partial_t n_{\rm ind}$ is then
\begin{equation}
\label{eq:j0}
{\bf j}_{\rm ind}^{(0)}({\bf r},t)=\frac{-e}{16 v^2}\frac{1}{r}\left(\hat z\times\frac{\partial^2{\bf U}(t-r/v)}{\partial t^2}\right),
\end{equation}
where we have used the boundary condition $j_{\rm ind}(r\to\infty)=0$.
The charge and current densities at time $t$ are determined by the external perturbation at a retarded time $t-r/v$, which reflects the relativistic nature of Dirac fermions (with $v$ as the ``speed of light'').
  
For concreteness, consider a single impurity whose spin is pointing along $\hat x$ in equilibrium. 
Under a constant magnetic field ${\bf B}=B_0\hat y$, the impurity magnetic moment precesses and ${\bf U}(t)=\lambda_{||} [\cos(\gamma B_0 t)\hat x+\sin(\gamma B_0 t)\hat z]$, where $\gamma$ is the gyromagnetic ratio and we have neglected magnetic damping. Using $\gamma\simeq 1.76\times 10^{11}{\rm Hz}/{\rm T}$, $v=5\times10^5 {\rm m/s}$ and\cite{garate2009} $\lambda_{||}\simeq 50 {\rm meV nm^2}$, Eqs.~(\ref{eq:n0}) and (\ref{eq:j0}) read 
\begin{eqnarray}
\label{eq:numbers}
n_{\rm ind}^{(0)}\left[\frac{{\rm e}}{\mu m^2}\right] &=& 3.3\frac{B_0[{\rm T}]}{r[{\rm nm}]^2}\left[\sin\theta(t^*)+\frac{\gamma B_0 r}{v}\cos\theta(t^*)\right] ({\hat r}\cdot\hat y)\nonumber\\ 
{\bf j}_{\rm ind}^{(0)}\left[\frac{{\rm \mu A}}{{\rm m}}\right]&=&94 \frac{B_0[{\rm T}]^2}{r[\rm nm]}\cos\theta(t^*)\,\hat y,
\end{eqnarray} 
where $t^*=t-r/v$ and $\theta(t)=\gamma B_0 t$. 
$n_{\rm ind}^{(0)}$ is large enough to be measurable by a field effect transistor located in the vicinity (e.g. $r\simeq 10 {\rm nm}$) of the precessing local moment.
Likewise ${\bf j}_{\rm ind}^{(0)}$ may be experimentally accessible through its associated magnetic field ($\propto 1/r$ for large $r$), to be distinguished from the dipolar field ($\propto 1/r^3$) originating from the impurity moment.
The ${\rm AC}$ characteristics of ${\bf j}_{\rm ind}$ distinguish this effect qualitatively from the charge pumping phenomenon that occurs in more ordinary ferromagnet/paramagnet interfaces\cite{wang2006}. 

Next, we generalize Eqs.~(\ref{eq:n0}) and (\ref{eq:j0}) to $k_F\neq 0$.
Linearizing Eq.~(\ref{eq:f_0}) in $\Omega$ (which is akin to neglecting ``retardation effects'') and assuming $k_F r\gg 1$ we arrive at $n_{\rm ind}=n_{\rm ind}^{(0)}+\Delta n_{\rm ind}$ and $j_{\rm ind}=j_{\rm ind}^{(0)}+\Delta j_{\rm ind}$, where
\begin{eqnarray}
\label{eq:dn}
\Delta n_{\rm ind}&\simeq&-\frac{e}{4 \pi^2 v^2}(\hat r\times\hat z)\cdot\frac{\partial{\bf U}}{\partial t}\frac{k_{F}}{r}\nonumber\\
\Delta j_{\rm ind}&\simeq&\frac{e k_F}{4 \pi^2 v^2}\left(\hat z\times\frac{\partial^2{\bf U}}{\partial t^2}\right)\log r+{\rm const}.
\end{eqnarray}
In Eq.~(\ref{eq:dn}) we have neglected terms $\propto\cos(2 k_F r),\sin(2 k_F r)$ because their amplitudes are small.
Hence, when $k_F r\gg1$, $n_{\rm ind}$ and ${\bf j}_{\rm ind}$ decay more slowly with the distance from the local moment than when $k_F=0$.

{\em Conclusions.---}
Our analytical expressions for the coupled spin-charge response functions in a ballistic helical metal underlie a host of novel and potentially observable phenomena that arise when magnetic impurities or patterned magnetic films are deposited on the surface of a TI. Applied gate voltages and transport currents lead to unconventional RKKY interactions between two such impurities or magnetic monolayers, and oscillatory charge currents are generated by magnetic moments precessing under an external magnetic field.
Similar effects should also occur in ordinary 2DES with Rashba SO interaction, though in a much weaker form because the Rashba SO strength is typically much smaller than the Fermi energy.
It would be interesting to replicate our calculation for a diffusive helical metal, comparing it with Ref.~[\onlinecite{burkov2004}]. 

We acknowledge helpful interactions with I. Affleck, J. Folk,  H.-M. Guo and H. Karimi, as well as financial support from NSERC and CIfAR.
I.G. is a CIfAR Junior Fellow.


\begin{thebibliography}{30}
\bibitem{ti} M.Z. Hasan and C.L. Kane, arXiv:1002.3895 (2010) and references therein.
\bibitem{qi2008} X.-L. Qi {\em et al.}, Phys. Rev. B {\bf 78}, 195424 (2008).
\bibitem{yokoyama2009} T. Yokoyama, Y. Tanaka and N. Nagaosa, Phys. Rev. B {\bf 81}, 121401 (2010).
\bibitem{garate2009} I. Garate and M. Franz, arXiv:0911.0106v1 (2009).
\bibitem{ye2010} F. Ye {\em et al.}, arXiv:1002.0111v1 (2010).
\bibitem{gao2009} J. Gao {\em et al.}, Phys. Rev. B {\bf 80}, 241302(R) (2009).
\bibitem{biswas2009} R. Biswas and A.V. Balatsky, arXiv:0910.4604v1 (2009).
\bibitem{liu2009} Q. Liu {\em et al.}, Phys. Rev. Lett. {\bf 102}, 156603 (2009).
\bibitem{huang2006} W.-M. Huang, C.-H. Chang and H.-H. Lin, Phys. Rev. B {\bf 73}, 241307(R) (2006).
\bibitem{graphene} B. Wunsch {\em et al.}, New J. Phys. {\bf 8}, 318 (2006); Y. Barlas {\em et al.}, Phys. Rev. Lett. {\bf 98}, 236601 (2007); E.H. Hwang and S. Das Sarma, Phys. Rev. B {\bf 75}, 205418 (2007). 
\bibitem{barlas2007} Appendix A of Y. Barlas, PhD Thesis, University of Texas at Austin, 2008. 
\bibitem{imamura2004} H. Imamura, P. Bruno and Y. Utsumi, Phys. Rev. B {\bf 69}, 121303(R) (2004).
\bibitem{caveat} This statement applies to low-energy universal behavior. In practice there is a paramagnetic contribution coming from high energies, where the energy dispersion of the surface states is no longer linear in momentum.
\bibitem{rkky} See e.g. C. Kittel, Solid State Phys. Adv. Res. Appl. {\bf 22}, 1 (1968).
\bibitem{saremi2007} S. Saremi, Phys. Rev. B {\bf 76}, 184430 (2007).
\bibitem{brey2007} L. Brey, H.A. Fertig and S. Das Sarma, Phys. Rev. Lett. {\bf 99}, 116802 (2007).
\bibitem{baltensperger1990} W. Baltensperger and J.S. Helman, Appl. Phys. Lett. {\bf 57} (27), 2954 (1990).
\bibitem{dm} I. Dzyaloshinskii, J. Phys. Chem Solids {\bf 4}, 241 (1958); T. Moriya, Phys. Rev. {\bf 120}, 91 (1960).
\bibitem{neq} N.F. Schwabe, R.J. Elliot and N.S. Wingreen, Phys. Rev. B {\bf 54}, 12953 (1996); V.I. Kozub and V. Vinokur, Appl. Phys. Lett. {\bf 87}, 062507 (2005); P.M. Haney, C. Heiliger and M.D. Stiles, Phys. Rev. B {\bf 79}, 054405 (2009).
\bibitem{wang2006} X. Wang {\em et al.}, Phys. Rev. Lett. {\bf 97}, 216602 (2006).
\bibitem{burkov2004} A.A. Burkov, A.S. Nunez and A.H. MacDonald, Phys. Rev. B {\bf 70}, 155308 (2004).
\end{thebibliography}
\end{document}